# Adaptive Alert Throttling for Intrusion Detection Systems

Gianni Tedesco (Corresponding Author)43 Runswick Terrace
Bradford
BD5 8LR
United Kingdom
+44 7811 328 171
gianni@scaramanga.co.uk

Uwe Aickelin
School of Informatics
University of Bradford
Bradford BD7 1DP
UK
+44 1274 233609
u.aickelin@bradford.ac.uk

**Abstract**

Each time that an intrusion detection system raises an alert it must make some attempt to communicate the information to an operator. This communication channel can easily become the target of a denial of service attack because, like all communication channels, it has a fixed capacity. If this channel can become overwhelmed with bogus data, an attacker can quickly achieve complete neutralisation of intrusion detection capability. Although these types of attack are very hard to stop completely, our aim is to present techniques that improve alert throughput and capacity to such an extent that the resources required to successfully mount the attack become prohibitive.

## 1 Introduction

As global awareness of information security issues has increased, so has the proliferation of intrusion detection technology. It is widely agreed that network intrusion detection systems (NIDS or IDS) are quickly becoming a crucial part of Internet security infrastructure. In March 2001 there was a media furore [ZDNet UK, 2001] when the FBI Internet crime division issued a warning concerning the then unreleased Stick [Coretez, 2002] program which "essentially disarms intrusion detection systems." With attackers now focusing their energy on network intrusion detection systems, the importance of protecting them becomes ever greater.

IDSs come in many different forms [Escamilla, 1998]. Our work is based on network IDSs, which detect attacks by directly analysing network traffic in real-time. Techniques used by NIDSs still have a lot of room to evolve. Ning [2001] identifies a number of problems associated with current systems. Examples of current NIDS research include methods of identifying new or varied attacks[Ning, 2001], 'honeypots' [Hogwash, 2003] and anomaly detection [SPADE, 2001]. Our area of research focusses on preventing denial of service attacks.

There are, in fact, numerous types of denial of service attack possible against an intrusion detection system [Ptacek, 1998], but we will focus on one particular attack which is fundamental to all systems: alert flooding. We will also focus primarily on network intrusion detection systems using the signature matching model described below. However, most of the techniques presented will be generally applicable to any type of intrusion detection system.

The pattern matching[Cisco, 2002] model is currently the most commonly used methodology for detecting intrusion attempts. In this model the IDS is configured with a database of known attack patterns (called signatures). An example of a signature may look something like Figure 1. This signature alerts on traffic generated by the well-known "BackOrifice" trojan horse program and detects any incoming packets destined to user datagram protocol (UDP) port 31337, containing a specific sequence of bytes anywhere within its payload.

Figure 1: A Sample Snort Rule

```
alert udp $EXTERNAL_NET any -> $HOME_NET 31337 (msg:"BACKDOOR BackOrifice access";
content: "|ce63 d1d2 16e7 13cf 39a5 a586|";)
```

Alert flooding attacks are achieved by transmitting packets that simulate intrusion attempts and which the IDS will recognise as such. Taking the example signature in Figure 1, an attacker must craft a UDP packet, set the destination port to 31337, include the sequence of bytes given in the signature and flood the target network with these packets.

The possible ramifications of this type of attack against an intrusion detection system are threefold:

1. Sensor storage becomes full, preventing further logging.

2. Sensor exceeds maximum alert throughput, causing alerts to be lost, or the sensor to cease functioning.

3. The analyst becomes deluged with false information and becomes unable to distinguish real attacks from the false ones.

The alert flooding technique has been automated, and hence popularised, by tools such as Stick [Coretez, 2002] and Snot [Sniph, 2001] which read in signatures directly from the freely available Snort [Roesch, 1999] IDS. Each packet sent could also have crucial fields such as source and destination address modulated by adding random data into them. This makes it difficult to block the attack using a simple packet filter or firewall.

Alert floods can also be exacerbated by the poor alerting performance of IDS systems in general. A quick examination of the Snort [Roesch, 1999] system reveals that, in its preferred output mode (called "unified"), Snort flushes its buffers needlessly in at least two places. This causes reduction in the effectiveness of the buffering and on UNIX like systems results in added system call overhead for every logged alert.

Performance in this area can be understandably overlooked by the IDS system designer. After all, good engineering practise tells us to optimise for the common case, and in the world of intrusion detection an alert usually is not the common case. In fact, on a high speed network it should be a very rare event indeed.

The next section of this paper will show that previous approaches have focused on removing short cuts to performing the attack in order to raise the cost to the attacker. In section 3 we define the problem more carefully and examine the flaws in existing solutions. We then take the line of thought further, suggesting how an adaptive throttling technique can increase the cost to the attacker as the magnitude of the attack increases. We will also provide some experimental data, and some calculations of the effectiveness of our technique. We then finish by presenting a summary and some concluding remarks.

## 2 Current State of the Art

The Snort [Roesch, 1999] team addressed the problems of wide spread proliferation of automated alert flooding tools like Stick [Coretez, 2002] and Snot [Sniph, 2001] in their 1.8 release. Their solution was to implement a Transmission Control Protocol (TCP) state tracking system which they called "stream4".

TCP is the transport protocol of the majority of Internet traffic. It is a connection oriented two-way stream protocol (similar to a telephone call) and has a connection initialisation procedure known as the "three-way handshake". A TCP message (called a segment) cannot exist outside of an established connection in much the same way that a telephone conversation cannot exist without first dialling a number and waiting for the other side to pick up the receiver.

By keeping track of TCP connection states, stream4 is able to ignore any segments which are not part of such a conversation. In order to make the IDS raise an alert the attacker is now forced to transmit at least three segments, rather than just one. More importantly, because the three-way handshake requires two hosts to be communicating, the external attacker must find a host on the monitored network willing to participate.

Currently most systems keep track of TCP state. This is mainly to protect against desynchronisation attacks as described by Ptacek and Newsham [Ptacek, 1998], but there is also the additional benefit of making sure that there is no such short cut in carrying out an alert flooding attack. Further to performing TCP state tracking, it is also possible to track any application layer state, enabling us to remove short cuts even for protocols running over stateless transports such as UDP.

Looked at it pragmatically we can assert that with the stateless implementation the attacker had a short cut to make the attack cheaper, equivalent to simply picking up a telephone receiver and starting talking. Naturally, the solution is to remove that short cut. This forces the attacker to find a valid number, dial it, and wait for an answer.

While this approach is a positive step, it cannot cover all cases: for example some signatures must ignore state information due to the fact that some exploits can exist as a single packet, or because they work over inherently stateless protocols. We will go on to show ways to reduce the alerting throughput during an alert flood attack, which will work regardless of signature configuration.

## 3 Adaptive Data Reduction

We will present a technique for reducing alerting throughput during an alert flood attack by detecting that a flood is occurring and then adapting to the threat by making the IDS more terse, even if that means discarding or "dropping" alerts. We show that the technique scales up to effectively nullify the threat of an alert flood attack. A similar approach to this has been successfully deployed in order to drastically slow the spread of worms and viruses across the Internet [Twycross, 2003]. We then show how some simple compression techniques can be used to ensure that we can still log important information about the attack.

We have described that the problem is fundamentally that of resource exhaustion. For example our human IDS operator is a resource, and one which cannot cope with having to examine many thousands of bogus alerts at the rate at which a sustained attack can produce them.

There are two approaches to solving this type of problem: one is to increase the amount of resource you have, the other is to reduce the amount of resources required. While it is conceivable that one could scale the sensor hardware to be fully able to cope with alert floods at a given rate for a given length of time it seems rather more complex to scale the human operator.

Unfortunately, the processing capacity of human computer operators scales linearly at absolute best [Brooks, 1995]. However for complex analytical tasks requiring coordination and communication, adding more people is so inefficient that it can even make a task take longer. Besides, even if humans could scale to the task linearly it would seem an incredibly expensive proposition for an organisation to multiply the number of operators even by a factor of ten. It would seem that the more beneficial approach would be that of data reduction.

### 3.1 Token Bucket Filter

Perhaps the simplest way to reduce data output while maintaining the same intrusion detection capability is by making minor modifications to the signatures to make sure that the IDS is as terse as possible. Such modifications are often used to reduce the number of false positive alerts generated. In fact generally speaking, signatures are usually a subtle compromise between allowing false negative and false positive alerts.

One way to make the IDS less verbose is to fine-tune signatures to examine only those packets destined for the relevant hosts. Let us consider BIND, a DNS server software infamous for its security vulnerabilities. In this situation, the signatures may be modified to only look for BIND exploits if the destination address on the packet matches a pre-defined list of DNS servers. Of course, the operator may actually be interested to know that someone is attempting a BIND exploit on a workstation or a web server. That is to say, this approach tips the false alarm compromise towards the false negative side.

Perhaps a more beneficial approach would be adaptive. In the case where there is no need to conserve alert throughput we log all suspicious activity. Once a certain rate threshold has been met, indicating an alert flood attack, we could switch to a more terse behaviour or even drop alerts exceeding the threshold all together. This can be achieved very efficiently by using a token bucket filter [Turner, 1986].

A token bucket filter is an algorithm for controlling the rate of flow of data. Token bucket filters have traditionally been used in a number of applications where rate limiting has been needed. Some good examples are:

1. Network bandwidth management systems [Woodruff, 1988] .

2. Flood protection in network chat / text conferencing systems.

3. Flood protection for programs that log externally generated events.

4. Congestion control in network transport protocols.

A token bucket filter has two parameters, bucket size, and token rate. Tokens are generated at the token rate and stored in a buffer called the "bucket". If the bucket becomes full, the extra tokens are just discarded. Each alert that arrives must have a token to pass through the filter. Any alert that does not have a token is called "over-limit" and does not pass the filter. If the alert rate is less than the token-rate then credit is allowed to accumulate in the bucket. This stored credit allows for the alert-rate to temporarily exceed the token rate, which is called a "burst".

The token bucket filter could be applied per signature, per attack type, globally, or even in complex hierarchies as in HTB3 [Devera, 2002]. Firestorm [Tedesco, 2003] is one system that can do this and has configurable per signature and per "generator" (attack type) thresholds. The filtering is readily configurable as an extension to the Snort [Roesch, 1999] signature format which it supports natively. Firestorm completely drops all over-limit alerts.

We can perform a simple test with the Firestorm [Tedesco, 2003] system running off-line against a freely available Shmoo Group "CCTF" [Shmoo Group, 2001] data set which contains a high bandwidth Internet packet fragmentation attack constituting an alert flood. The capture file contains 587,482 packets transmitted in the space of 80 seconds, which are part of the attack.

The Firestorm system has been chosen as it already supports token bucket filtering and supports the same signatures and intrusion detection techniques as the popular Snort[Roesch, 1999] system. This makes it roughly representative of IDS systems that are currently deployed. It is also free software, with available source code that allows us to readily add instrumentation and measurement code.

In the both tests we have a full signature database loaded containing around 1,600 signatures, the network data is being read directly from the hard disk. The test machine was an Apple Mac 667 MHz PowerBook G4 running Linux 2.4. The results shown are an average of three runs. This to factor out any random fluctuations such as disk seek latency.

The first run (#1) is a control run. The second run (#2) is identical except for the addition of a token bucket filter. The filter applies to all IP-fragmentation related alerts and is configured with a token rate of 1 per second, and a bucket size of 10 tokens. The parameters are rather arbitrary but seem reasonable to differentiate a real alert flood from a small sequence of genuine attacks.

Table 1: Token Bucket Filter Alert Throughput

| # | Data Size (Kilo Bytes) | Alerts | CPU Time (seconds) | Elapsed Time |
|---|---|---|---|---|
| 1 | 472,192 | 300,299 | 35.05 | 81.341 |
| 2 | 172 | 109 | 29.09 | 50.258 |

As we can see in Table 1, the amount of data logged was reduced by several orders of magnitude and the run time decreased disproportionately to the CPU time. While the run time was reduced by 60% the CPU time only reduced by 20% . This indicates that with the token bucket filter enabled, the Firestorm process is not wasting as much time waiting for I/O completion.

From these results it is clear that we can effectively boost performance and capacity, allowing the sensor to carry on working during an alert flood rather than becoming overwhelmed and possibly exhausting the storage on the sensor. Even if the attack contained twice as many packets in the same space of time, it would not double the amount of data logged as the token rate is fixed.

## 3.2 Alert Compression

Dropping alerts entirely may seem a less than ideal way of dealing with an alert flooding situation. The IDS operator may be uncomfortable with not having all the evidence with which to make judgements. However if we assume that alert flood attacks are mainly repetitive, we exploit this similarity to achieve effective compression of over-limit alerts instead of dropping them.

If we assume that the attack is the same packet repeatedly then we can use a "run length encoding" (RLE) to represent it. RLE is a simple compression technique which replaces recurring sequences of symbols (called runs) with a single symbol and a run count **N**. To decompress one simply copies the symbol into the output stream **N** times. This is an approach familiar to UNIX users who have ever tried to flood the syslog program and seen its "last message repeated N times" warning.

To implement RLE compression in our case all that is required is to store a reference to the first over-limit alert against the token bucket data structure and increment a counter for all further over-limit alerts. When there is enough credit in the token bucket to permit new alerts, we flush out the alert and the counter to permanent storage.

The operator can then see that alert **A** was repeated **X** times. The only data lost is the timing of the **(X - 1)** alerts which constitute the run. However, the time can be encoded along with the alert as delta values. For example if the IDS timestamps are kept in millisecond resolution we could use a 16 bit unsigned integer to represent the time deltas. That would allow us to keep totally accurate timestamps provided that no two alerts in the run are more than 65 seconds apart.

An advantage of combining delta compression with RLE is the ability to present the operator with one composite alert for each run while still providing the ability to see the complete un-filtered data or "drill down" if that is required. This can be implemented by only displaying the first alert in each run and providing the operator with an option to see the whole run, which could be decompressed on the fly.

Using our previous experiment as a basis, we can calculate the storage requirements for this type of compression. The Firestorm alert log format has enough reserved space for an RLE run counter which means that RLE compression has exactly the same storage requirements as the previous "token bucket only" result. We also provide a comparison with popular compression program gzip [Gailly, 2003]. In this case we simply ran the gzip command with default parameters and took the resulting file size.

Table 2: Comparison of Compression Techniques

| Algorithm | Data Size (Kilo Bytes) | Compression |
|---|---|---|
| Uncompressed | 472,192.6 | 1 |
| RLE Only | 171.578 | 2,752.056 |
| RLE With Timestamp Delta | 754.857 | 625.539 |
| Gzip | 5,380.11 | 87.766 |

As Table 2 shows, the RLE combined with delta compression provides a very impressive compression ratio (almost an order of magnitude greater than existing techniques) and without loss of fidelity in the case where all the alerts are identical.

To extend this technique we can further generalise our assumption that alerts are identical. In reality an attacker can randomly modulate fields in the packet structures, notably source and destination addresses (cf. Stick [Coretez, 2002].) In this instance we can use the delta compression technique for those extra fields or indeed for *all* fields in our alert as opposed to just the timestamp.

If faced with modulation within the actual packet payload one can choose a trade-off between losing all payload data for alerts in the run or saving all data, possibly by using an algorithm like rsync [Tridgell, 2000] as a generic delta compressor.

A real world system could be configured and fine tuned allowing the operator to make the trade-off between efficiency and data fidelity based on operational parameters such as hardware and network capacity and the perceived threat of alert flooding. For example, the operator may decide that the IP type-of-service (TOS) field provides so little information that it is not worth logging.

## 4 Summary and Conclusions

Alert flooding is a problem that will probably always exist with intrusion detection systems and one that cannot be eliminated entirely. However, we have shown that it is possible to drastically reduce the effects by recognising an attack and responding proportionately.

We detect the alert flooding attack using a token bucket filter and react by either dropping or compressing the alerts which make up the flood. We showed that delta compression of relevant fields can provide massive compression ratios in this situation, meaning that we can minimise the amount of useful data that is lost in reacting to the threat.

As an added benefit of the techniques presented, we are also able to group the many related alerts that make up the attack together, greatly simplifying analysis.

However, more investigation is needed to produce optimal token bucket filter parameters. An ideal scheme would probably be hierarchical, providing great flexibility in configuration.

A method for efficiently compressing full packet payload without loss of fidelity would also be useful. This should not present a serious problem as the rsync [Tridgell, 2000] algorithm provides a good delta compression solution. There is also a substancial body of knowledge on generic data compression algorightms. Gzip [Gailly, 2003] would appear to be a good choice.

## Acknowledgements

We would like to thank Louisa Parry, Matthew Hall and John Leach for input with writing the article.

# References

[Brooks, 1995] Brooks, Frederick. P., "The Mythical Man-Month" *Addison Wesley.* 1995.

[Devera, 2002] Devera, Martin., "Hierarchical token bucket theory." 2002.

http://luxik.cdi.cz/ devik/qos/htb/manual/theory.htm

[Cisco, 2002] Cisco Systems, The Science of Intrusion Detection System Attack Identification. 2002. http://www.cisco.com/warp/public/cc/pd/sqsw/sqidsz/prodlit/idssa wp.htm

[Coretez, 2002] Coretez, G. "Fun with Packets: Designing a Stick." *Endeavor Systems, Inc.*

2002. [Escamilla, 1998] Escamilla, "Intrusion Detection", Wiley, ISBN: 0-471-29000-9. 1998.

[Gailly, 2003] Gailly, Jean-Loup., Adler, Mark., The gzip compression algorithm.

2003. http://www.gzip.org/

[Hogwash, 2003] Hogwash. 2003.

http://hogwash.sourceforge.net/.

[Ning, 2001] Ning, D.S. Reeves and Y. Cui, "Correlating Alerts Using Prerequisites of Intrusions", Department of Computer Science, North Carolina State University, 2001, Technical Report TR-2001-13.

[Ptacek, 1998] Ptacek, T. H., Newsham, N. N., "Insertion, Evasion and Denial of Service: Eluding Network Intrusion Detection." *Secure Networks Inc.,* January 1998.

[Roesch, 1999] Roesch, M., "Snort - Lightweight Intrusion Detection for Networks." *13th Systems Administration Conference, USENIX,* 1999.

http://www.snort.org/

[Shmoo Group, 2001] Shmoo Group CCTF Defcon Data.

2001 http://www.shmoo.com/cctf/

[Sniph, 2001] Sniph. Snot. 2001.

http://www.stolenshoes.net/sniph/index.html

[SPADE, 2001] Hoagland, James. Staniford, Stuart. Statistical Packet Anomaly Detection. Silicon Defense. 2001.

http://www.silicondefense.com/software/spice/.

[Tedesco, 2003] Tedesco, G., "Firestorm Network Intrusion Detection System."

2003. http://www.scaramanga.co.uk/firestorm/

[Tridgell, 2000] Tridgell, A., "Efficient Algorithms for Sorting and Synchronization." *The Australian National University,* April 2000.

[Turner, 1986] J. Turner, New directions in communications (or which way to the information age?), IEEE Communications 24, 815. 1986.

[Twycross, 2003] Jamie Twycross, Matthew M. Williamson, Implementing and testing a virus throttle, HP Labs report HPL-2003-103, http://www.hpl.hp.com/techreports/2003/HPL-2003-103.html, 2003.

[Woodruff, 1988] G. Woodruff, R. Rogers, and P. Richards, A congestion control framework for high-speed integrated packetized transport, IEEE Globecomm 88, 1988, pp. 7.1.11.1.5

[ZDNet UK, 2001] ZDNet UK News. 2001.
http://news.zdnet.co.uk/internet/security/0,39020375,2085099,00.htm

## Biographies

Gianni Tedesco has worked in the security industry for 3 years, working around the Leeds/Bradford area. He is the author of the Firestorm Network Intrusion Detection System which is a high-performance signature based IDS. He has also worked on other projects within the security arena such as the Linux Netfilter firewalling framework.

Uwe Aickelin is a Lecturer in Computer Science at the University of Bradford. Previously, he worked as a lecturer in Mathematical Sciences at the University of the West of England. His main research interests are Artificial Immune Systems, Intrusion Detection, Scheduling and Evolutionary Computation. Uwe holds a PhD and a Masters degree from the University of Wales, Swansea and an undergraduate degree from the University of Mannheim.